\documentclass[preprint2]{aastex62}

\graphicspath{{./}{}}

\shorttitle{Ionized Gas Properties of Galaxies in ZFIRE Proto-Clusters}
\shortauthors{Alcorn et al.}

\begin{document}

\title{A Tale of Two Clusters: An Analysis of Gas-Phase Metallicity and Nebular Gas Conditions in Proto-cluster Galaxies at $z\sim2$}

\newcommand{\myemail}{lyalcorn@tamu.edu}
\newcommand{\halpha}{H$\alpha$}
\newcommand{\hbeta}{H$\beta$}
\newcommand{\Hbeta}{H$\beta$}
\newcommand{\OIII}{[\hbox{{\rm O}\kern 0.1em{\sc iii}}]}
\newcommand{\NII}{[\hbox{{\rm N}\kern 0.1em{\sc ii}}]}
\newcommand{\OII}{[\hbox{{\rm O}\kern 0.1em{\sc ii}}]}
\newcommand{\SII}{[\hbox{{\rm S}\kern 0.1em{\sc ii}}]}
\newcommand{\zfire}{{\sc ZFIRE}}
\newcommand{\vrot}{$V_{rot}$}
\newcommand{\vtwo}{$V_{2.2}$}
\newcommand{\sint}{$\sigma_{\rm int}$}
\newcommand{\sigmag}{$\sigma_g$}
\newcommand{\Msun}{M$_{\odot}$}
\newcommand{\Mstar}{M$_{\star}$}
\newcommand{\Mdyn}{M$_{\rm vir}$}
\newcommand{\Mbar}{M$_{\rm baryon}$}
\newcommand{\kms}{km~s$^{-1}$}
\newcommand{\chisquare}{$\chi^2$}
\newcommand{\zsimtwo}{$z\sim2$}
\newcommand{\zltone}{$z\le1.7$}
\newcommand{\sof}{S$_{0.5}$}
\newcommand{\reducedchisquare}{$\chi^2_{\nu}$}
\newcommand{\voversig}{$V_{2.2}/\sigma_g$}
\newcommand{\sersic}{S{\'e}rsic}

\correspondingauthor{Leo Y. Alcorn}
\email{lyalcorn@tamu.edu}

\author[0000-0002-2250-8687]{Leo Y. Alcorn}
\affil{Department of Physics and Astronomy, Texas A\&M University, College Station, TX, 77843-4242 USA}
\affil{George P.\ and Cynthia Woods Mitchell Institute for Fundamental Physics and Astronomy, Texas A\&M University, College  Station, TX, 77843-4242}
\affiliation{LSSTC Data Science Fellow}

\author[0000-0002-8984-3666]{Anshu Gupta}
\affiliation{School of Physics, University of New South Wales, Sydney, NSW 2052, Australia}
\affiliation{ARC Centre of Excellence for All Sky Astrophysics in 3 Dimensions (ASTRO 3D), Australia}

\author[0000-0001-9208-2143]{Kim-Vy Tran}
\affiliation{Department of Physics and Astronomy, Texas A\&M University, College Station, TX, 77843-4242 USA}
\affiliation{George P.\ and Cynthia Woods Mitchell Institute for Fundamental Physics and Astronomy, Texas A\&M University, College  Station, TX, 77843-4242}
\affiliation{School of Physics, University of New South Wales, Sydney, NSW 2052, Australia}
\affiliation{ARC Centre of Excellence for All Sky Astrophysics in 3 Dimensions (ASTRO 3D), Australia}

\author[0000-0003-1362-9302]{Glenn G. Kacprzak}
\affiliation{Swinburne University of Technology, Hawthorn, VIC 3122, Australia}
\affiliation{ARC Centre of Excellence for All Sky Astrophysics in 3 Dimensions (ASTRO 3D), Australia}

\author[0000-0002-9211-3277]{Tiantian Yuan}
\affiliation{Swinburne University of Technology, Hawthorn, VIC 3122, Australia}
\affiliation{ARC Centre of Excellence for All Sky Astrophysics in 3 Dimensions (ASTRO 3D), Australia}

\author[0000-0003-1420-6037]{Jonathan Cohn}
\affil{Department of Physics and Astronomy, Texas A\&M University, College Station, TX, 77843-4242 USA}
\affil{George P.\ and Cynthia Woods Mitchell Institute for Fundamental Physics and Astronomy, Texas A\&M University, College  Station, TX, 77843-4242}

\author[0000-0001-6003-0541]{Ben Forrest}
\affiliation{Department of Physics \& Astronomy, University of California, Riverside, 900 University Avenue, Riverside, CA 92521, USA}

\author[0000-0002-3254-9044]{Karl Glazebrook}
\affiliation{Swinburne University of Technology, Hawthorn, VIC 3122, Australia}

\author{Anishya Harshan}
\affiliation{School of Physics, University of New South Wales, Sydney, NSW 2052, Australia}

\author[0000-0001-8152-3943]{Lisa J. Kewley}
\affiliation{Research School of Astronomy and Astrophysics, The Australian National University, Cotter Road, Weston Creek,
ACT 2611, Australia}
\affiliation{ARC Centre of Excellence for All Sky Astrophysics in 3 Dimensions (ASTRO 3D), Australia}

\author[0000-0002-2057-5376]{Ivo Labb\'e}
\affiliation{Swinburne University of Technology, Hawthorn, VIC 3122, Australia}

\author[0000-0003-2804-0648]{Themiya Nanayakkara}
\affiliation{Leiden Observatory, Leiden University, P.O. Box 9513, NL 2300 RA Leiden, The Netherlands}

\author[0000-0001-7503-8482]{Casey Papovich}
\affiliation{Department of Physics and Astronomy, Texas A\&M University, College Station, TX, 77843-4242 USA}
\affiliation{George P.\ and Cynthia Woods Mitchell Institute for Fundamental Physics and Astronomy, Texas A\&M University, College  Station, TX, 77843-4242}

\author[0000-0001-5185-9876]{Lee R. Spitler}
\affiliation{Department of Physics and Astronomy, Faculty of Science and Engineering, Macquarie University, Sydney, NSW 2109, Australia}

\author[0000-0001-5937-4590]{Caroline M. S. Straatman}
\affiliation{Sterrenkundig Observatorium, Universiteit Gent, Krijgslaan 281 S9, 9000 Gent, Belgium}

\begin{abstract}

The ZFIRE survey has spectroscopically confirmed two proto-clusters using the MOSFIRE instrument on Keck 1: one at $z=2.095$ in COSMOS and another at $z=1.62$ in UDS. 
Here we use an updated ZFIRE dataset to derive the properties of ionized gas regions of proto-cluster galaxies by extracting fluxes from emission lines \hbeta\ 4861\AA, \OIII\ 5007\AA, \halpha\ 6563\AA, \NII\ 6585\AA, and \SII\ 6716,6731\AA. 
We measure gas-phase metallicity of members in both proto-clusters using two indicators, including a strong-line indicator relatively independent of ionization parameter and electron density. 
Proto-cluster and field galaxies in both UDS and COSMOS lie on the same Mass-Metallicity Relation with both metallicity indicators. 
We compare our results to recent IllustrisTNG results, which reports no significant gas-phase metallicity offset between proto-cluster and field galaxies until $z=1.5$.
This is in agreement with our observed metallicities, where no offset is measured between proto-cluster and field populations.
We measure tentative evidence from stacked spectra that UDS high mass proto-cluster and field galaxies have differing \OIII/\Hbeta\ ratios, however these results are dependent on the sample size of the high mass stacks.

\end{abstract}

\keywords{galaxies -- evolution}

\section{Introduction} \label{sec:intro}

Proto-clusters are the precursor to the most extreme environments in the universe.
In the local universe ($z=0$), there is a relationship between environmental density and the properties of galaxies, i.e. morphology, stellar populations, and star-formation rate.
In particular, studies have found elevated gas-phase metallicity in local cluster galaxies \citep{Cooper2008,Ellison2009}, lower star-formation rates (SFR) in cluster galaxies \citep{Lewis2002,Grootes2017,Jarrett2017}, a greater fraction of elliptical and lenticular galaxies in denser environments \citep{Houghton2015}, redder colors with increasing density \citep{Blanton2005}, and a higher fraction of slow rotating galaxies in dense environments \citep{Cappellari2011}.
Since cluster environments show these correlations at low redshift ($z<0.5$), we wish to observe when the environment-dependent evolution of galaxies unfolds.
At higher redshifts, the number fraction of low SFR and quenched galaxies in denser environments decreases \citep{Nantais2013,Lee2015,Kawinwanichakij2017,Pintos-Castro2019}, until the star formation rate density of the universe reaches its peak at \zsimtwo\ \citep{Hopkins2006,Madau2014}.

The period known as ``Cosmic Noon" ($1.5<z<2.5$) is of particular interest as this is the stage where massive galaxies within proto-clusters are rapidly building their stellar populations \citep{Tran2010,Hatch2011,Koyama2013a,Lee2015,Shimakawa2018}. 
Analyses of metallicity \citep{Kacprzak2015,Tran2015, Kacprzak2016}, star-formation \citep{Koyama2013a,Koyama2013,Tran2015, Tran2017}, ionized gas characteristics \citep{Kewley2016}, gas content \citep{Koyama2017,Darvish2018}, and kinematics \citep{Alcorn2016, Alcorn2018} of proto-cluster members at $z>1.5$ showed no differences compared to field galaxies at the same epoch.
Some studies of galaxy sizes in proto-clusters vs. field observe an increase in proto-cluster star-forming galaxy sizes relative to the field \citep{Allen2015, Ito2019}, but others do not observe a difference \citep{Rettura2010,Bassett2013, Suzuki2019}.
Clear signatures of the morphology and color relation in dense environments have been observed in \citet{Cooper2006,Gobat2011,Papovich2012, Bassett2013,Strazzullo2016}. Additionally, an increase in merger rates in \zsimtwo\ clusters were observed \citep{Lotz2013, Watson2019}.

Cosmological simulations have found small ($0.05-0.1$ dex) metallicity enhancements in proto-clusters at $z>1.5$.
The simulations by \citet{Dave2011} determined a metallicity enhancement in galaxies in dense environments on the order of $\sim 0.05$ dex.
\citet{Kacprzak2015} performed hydrodynamical simulations with Gadget-3 \citep{Kobayashi2007}, determining that proto-cluster galaxies at \zsimtwo\ would be enhanced by $\sim$0.1 dex from field galaxies in the Mass-Metallicty Relation (MZR).
\citet{Gupta2018} observed that within the IllustrisTNG (100 Mpc)$^3$ simulation (TNG100), galaxies in proto-clusters show a $\sim$0.05 dex gas-phase metallicity enhancement compared to field galaxies starting at $z=1.5$, but no significant offsets at $z=2$ and higher. 
This possibly identifies the redshift when the cluster environment starts influencing member galaxies in terms of their gas-phase metallicities.
Here, we provide an observational counterpart to the \citet{Gupta2018} IllustrisTNG results, and determine the gas-phase metallicity offset between two proto-cluster populations and field galaxies at z=1.6 and 2.095.

The \zfire\ survey \citep{Tran2015, Nanayakkara2016} spectroscopically confirmed the over-dense region in the COSMOS field at $z=2.095$ \citep{Yuan2014}, and the over-dense region in the UDS field at $z=1.62$ \citep{Papovich2010, Tran2015}.
With the ZFIRE dataset of \zsimtwo\ proto-cluster galaxies, it is possible to extend kinematic and metallicity scaling relations to the proto-cluster environment at \zsimtwo. 
The ZFIRE dataset allows us to examine the interplay of the intra-cluster medium, galactic winds, and the interstellar medium (ISM) in a cluster and its member galaxies.
The proto-clusters observed in this study were originally identified using deep NIR imaging in \citet{Papovich2010, Spitler2012}.
Using rest-frame optical emission lines from members of the proto-cluster, we can directly compare nebular gas properties of cluster and field populations at this epoch.

The \zfire\ collaboration has increased its sample since the initial data release in both the UDS and COSMOS fields, and measured more nebular emission lines (\hbeta, \OIII) from galaxies in the previous sample \citep{Tran2015, Nanayakkara2016}.
The extended sample allows us to perform a re-analysis (originally by \citet{Kewley2016}) of the Baldwin-Phillips-Terlevich (BPT) Diagram \citep{Baldwin1981} on the basis of redshift, inferred stellar mass, and environmental density.
The sample of COSMOS galaxies with all BPT diagnostic emission lines is doubled (from 37 galaxies to 74), and 41 UDS galaxies are also analyzed using the BPT Diagram. 
Additionally, we stack spectra, not seen in the previous BPT \citep{Kewley2016} or MZR \citep{Tran2015, Kacprzak2015} studies.
This diagnostic will provide information on the nebular gas in each proto-cluster to determine if star-forming conditions evolve with redshift.

By including recent improvements in gas-phase metallicity indicators, we will re-examine galaxy metallicity and directly compare two proto-clusters at \zsimtwo\ to the field.
Gas-phase metallicity values for the same galaxy can differ between metallicity indicators by up to 1 order of magnitude \citep{Kewley2008}.
Discrepancies between different strong-line indicators are possibly the result of calibration on local HII regions, which depend on the measurement of electron temperature \citep{Pilyugin2005,Pilyugin2011,Maiolino2019}.
Calibrations using local HII regions can be flawed when used for high-redshift measurements because possible changes in the nebular gas properties of high-redshift galaxies have been observed compared to local galaxies.
In particular, galaxies at high redshift have been observed to have higher ionization parameter and electron density than local galaxies \citep{Brinchmann2008,Liu2008,Lehnert2009,Newman2012,Kewley2013,Tacconi2013,Newman2014, Shirazi2014, Steidel2014, Shapley2015,Steidel2016, Kaasinen2018, Harshan2019}.

We utilize the strong-line gas-phase metallicity diagnostics introduced in \citet{Pettini2004} (referred to herein as PP04) and the newer indicator by \citet{Dopita2016} (referred to as D16).
The PP04 indicator is widely used in high-redshift extragalactic astronomy \citep{Wuyts2014,Zahid2014,Sanders2015, Sanders2018} and is not sensitive to reddening (because it utilizes a ratio of \halpha\ to \NII).
However it is subject to the calibration issues noted above and is sensitive to the ionization parameter and electron density of a host galaxy, as well as the presence of shock-heated gas and low-level AGN.
The D16 indicator attempts to correct this by including the \SII\ doublet in this ratio, and photoionization models show that it is insensitive to ionization parameter and electron density.
We measure gas-phase metallicity using both strong-line indicators in this study to determine the effectiveness of the D16 indicator.

In Section \ref{sec:data} we review our data from Magellan (FOURSTAR), and Keck (MOSFIRE) and extensive public multi-wavelength observations. 
We discuss our line-fitting and flux extraction methods in Section \ref{sec:methods} including both individual objects (Section \ref{sec:emlines}) and stacked emission lines (Section \ref{sec:stacking}).
Our results for gas-phase metallicity between proto-cluster and field in UDS and COSMOS are shown in Section \ref{sec:metal}.
The nebular gas analysis of individual galaxies and our stacked objects using the BPT diagram is located in Section \ref{sec:bptsection}.
Finally, our results are put into the context of galaxy evolution in the proto-cluster environment in Section \ref{sec:analysis}.

In this work, we assume a flat $\Lambda$CDM cosmology with $\Omega_{M}$=0.3, $\Omega_{\Lambda}$=0.7, and H$_{0}$=70. 
At the COSMOS proto-cluster redshift, $z = 2.095$, one arcsecond corresponds to a proper angular scale of 8.33 kpc.
At the UDS proto-cluster redshift, $z = 1.62$, one arcsecond corresponds to a proper angular scale of 8.47 kpc.

\section{Data} \label{sec:data}
\subsection{Sample Selection} \label{sec:sample}
Our proto-cluster and field samples are drawn from the \zfire\ survey, a spectroscopic follow-up of ZFOURGE observations \citep{Straatman2016}.
ZFOURGE combines broad-band imaging in $K_s$ and the medium-band J$_{1}$, J$_{2}$, J$_{3}$, H$_{s}$, and H$_{l}$ filters to select objects using $K_s$-band images with an 80\% completeness limit of 25.5 and 25.8 AB magnitudes in the COSMOS and UDS fields, respectively\footnote{http://zfourge.tamu.edu/}.
Rest-frame UVJ colors are used to identify star-forming galaxies (SFGs), which should have prominent nebular emission lines.
We account for AGN using the \citet{Cowley2016} AGN catalog.

ZFOURGE uses FAST \citep{Kriek2009} to fit \citet{Bruzual2003} stellar population synthesis models to the galaxy spectral energy distributions and estimate observed galaxy properties. 
We calculated stellar masses with FAST using the spectroscopic redshift from MOSFIRE \citep{Nanayakkara2016}.
We assume a \citet{Chabrier2003} initial mass function with constant solar metallicity and an exponentially declining star formation rate, and a \citet{Calzetti2000} dust law. 
For a full summary, see \citet{Tran2015, Nanayakkara2016,Straatman2016}.

\subsubsection{UDS} \label{sec:udssample}

The UDS proto-cluster at $z=1.62$ was identified by \citet{Papovich2010} and \citet{Tanaka2010}, from the \citet{Williams2009} catalog of UDS, a subset of the UKIRT survey \citep{Lawrence2007}.
A ZFIRE analysis of the UDS proto-cluster identified 26 confirmed proto-cluster members between $1.6118 <z< 1.6348$, and 36 field galaxies at equivalent redshifts \citep{Tran2015}.
This proto-cluster was shown to have a strikingly low velocity dispersion of $\sigma_{cl}=254\pm50$ \kms\ \citep{Tran2015}, and notably showed an increase in star-formation with local density \citep{Tran2010}.
Previous studies of this proto-cluster showed no enhancement in gas-phase metallicity (using the PP04 indicator, commonly referred to as N2 in literature) or attenuation compared to field samples \citep{Tran2015}. 

\subsubsection{COSMOS} \label{sec:cosmossample}

The COSMOS proto-cluster was initially identified in \citet{Spitler2012} using photometric redshifts from ZFOURGE and confirmed with spectroscopic redshifts from MOSFIRE \citep{Yuan2014}.
This over-density consists of four merging groups, and is projected to evolve into a Virgo-like cluster at $z=0$ \citep{Yuan2014}.
Fifty-seven cluster members are identified to spectroscopic redshifts within $2.08<z<2.12$, and the cluster velocity dispersion was shown to be $\sigma_{cl}=552\pm52$ \kms.
This proto-cluster shows no enhancement in PP04 gas-phase metallicity compared to field samples \citep{Kacprzak2015}.
One hundred twenty-three star-forming field galaxies are selected from redshifts of $2.0<z<2.5$ from ZFIRE photometric modeling.

\subsection{MOSFIRE NIR Spectroscopy} \label{sec:mosfirespectra}

Observations were taken in December 2013, February 2014, January 2016, and February 2017 in the H and K filters covering 1.47-1.81 $\mu$m and 1.93-2.45 $\mu$m, respectively.
Seeing varied from  $\sim0.4\arcsec$ to  $\sim1.3\arcsec$ over the course of our observations.
Galaxies (star-forming, dusty, and quiescent) in over-dense regions in COSMOS and UDS were prioritized for observation over star-forming field galaxies.
For further information on target selection, see \citet{Nanayakkara2016}.

The spectra are flat-fielded, wavelength calibrated, and sky subtracted using the MOSFIRE data reduction pipeline (DRP)\footnote{http://keck-datareductionpipelines.github.io/MosfireDRP}. 
A custom ZFIRE pipeline corrected for telluric absorption and performed a spectrophotometric flux calibration using a type A0V standard star.
We flux calibrate our objects to the continuum of the standard star, and use ZFOURGE photometry as an anchor to correct offsets between photometric and spectroscopic magnitudes. 
The final result of the DRP are flux-calibrated 2D spectra and 2D 1$\sigma$ images used for error analysis with a flux calibration error of $<$10\% ($\sim$0.08 mag). 
For more information on ZFIRE spectroscopic data reduction and spectrophotometric calibrations, see \citet{Nanayakkara2016}.

\section{Methods} \label{sec:methods}
\subsection{Emission Line Flux Measurements} \label{sec:emlines}

\begin{figure*}[t!]
   \gridline{\fig{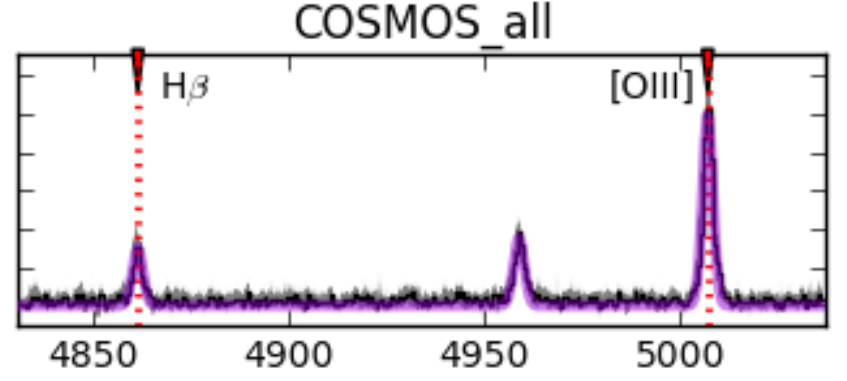}{0.5\textwidth}{}
          \fig{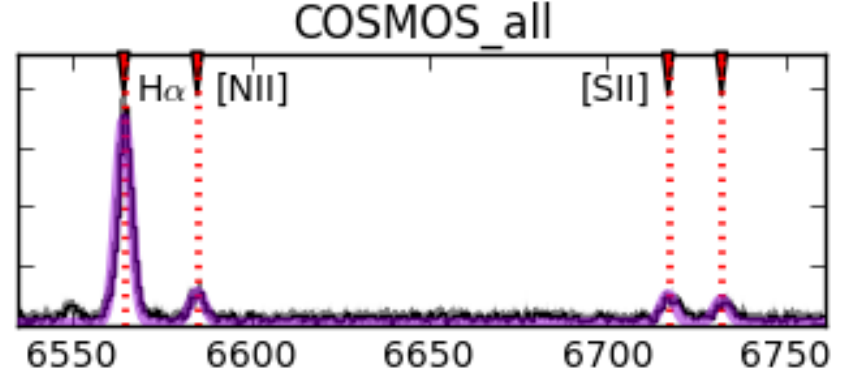}{0.5\textwidth}{}}
   \gridline{\fig{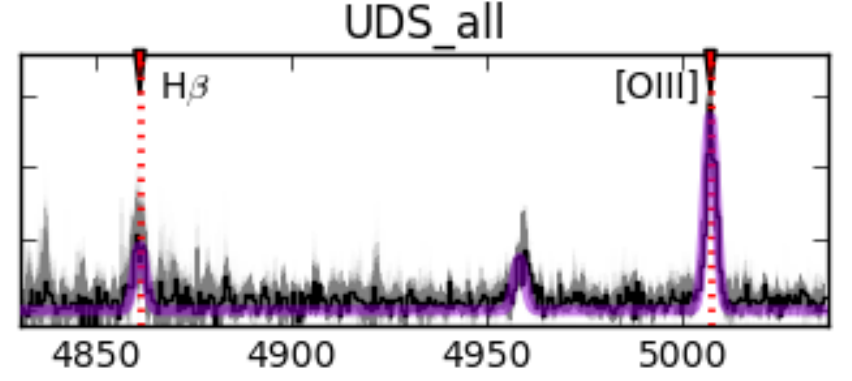}{0.5\textwidth}{}
          \fig{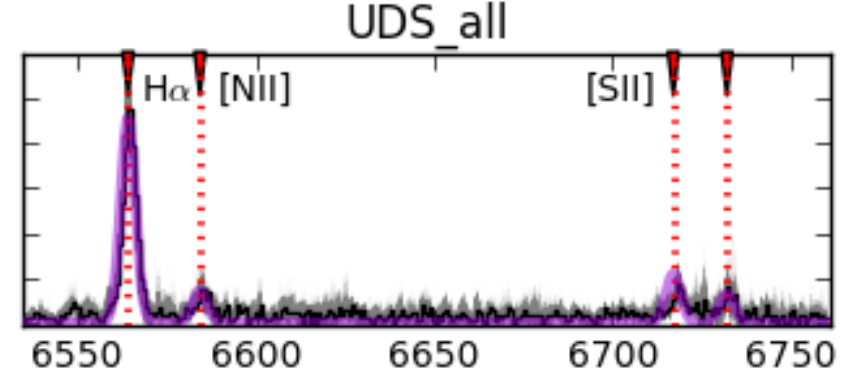}{0.5\textwidth}{}}
    
	\caption{Stacked rest-frame spectra of our sample, showing \halpha, \NII, \SII, \hbeta, and \OIII. 
	Black are the stacked spectra, which have been outlier-rejected. 
	Pink is the best-fit Gaussian profiles for the measured emission lines. 
	We show the individual bootstrapped spectra used for determining errors in grey.
	Red arrows at the top of each spectrum point to the measured emission lines, . 
	Left Column: The stacked \OIII\ and \hbeta\ emission lines. 
	In COSMOS, these lines are observed in the MOSFIRE H band. In UDS, these lines are observed in the MOSFIRE J band.
	Right: The stacked \halpha, \NII, and \SII\ emission lines. 
	In the COSMOS sample, these lines are observed in the K band, and in the UDS sample in the H band.
	The Gaussian fits are used to measure stacked flux ratios.}
\end{figure*}

We obtain emission line fluxes from our telluric-corrected and flux-calibrated spectra by fitting Gaussian profiles to emission lines.
Our 2D spectra are collapsed to a 1D spectrum by fitting a Gaussian to the spatial line profile, and summing along an aperture where each edge is twice the FWHM away from the center of the spatial profile of the emission line.
We then fit a Gaussian profile to the 1D spectrum of the emission line.
In Figure 1, it is clear that the \SII\ lines are well resolved and not blended; therefore we fit two single Gaussian profiles simultaneously at a fixed distance between their centroids.

\subsubsection{Measuring profiles} \label{sec:measuringprofiles}

From the best fit profile to the emission line, the standard deviation of the Gaussian is used to determine the range over which the line will be integrated, and we sum under the best-fit Gaussian from $-3\sigma < \lambda_{obs} <3\sigma$, where $\sigma=FWHM/2.35482$.
This value is our extracted flux measurement.
We measure noise by summing the error spectrum in quadrature over the same bounds as the signal.
For faint emission lines (\NII, \SII, \hbeta), we fit the line holding the Gaussian $\sigma$ fixed to the $\sigma$ measurement of the \halpha\ line (for \NII\ and \SII) or the \OIII\ emission line (for \hbeta).
If the emission line of an object is less than $2\times$ the measured noise in its region, we mark this object as an upper limit of the signal (unfilled points on Figure 2) and use the $1\sigma$ noise limit as our flux detection.

\subsubsection{Contamination By Sky Lines} \label{sec:skylines}

Due to strong sky emission line interference, we extracted only one UDS galaxy and 13 COSMOS galaxies with \halpha, \NII, and \SII\ unaffected by significant sky noise. However, 13 UDS galaxies and 47 COSMOS galaxies were extracted with upper limits due to sky emission.
Galaxies with \halpha, \NII, and \SII\ fluxes were used for our gas-phase metallicity analysis in Section \ref{sec:metal}.
For our flux ratio analysis (Section \ref{sec:bptsection}), we extracted fluxes for 26 COSMOS galaxies with \Hbeta, \OIII, \halpha, and \NII\ emission lines (15 of which had no emission lines with sky interference and 11 of which had at least one emission line with sky interference).
We also extracted fluxes for 11 UDS galaxies with all emission lines (5 with no sky interference, 6 with at least one emission line with sky interference).
We rejected objects with emission line SNR $<2$, which we discuss further in Section \ref{sec:indicatorcomp}.

\begin{figure*}[t] \label{fig:mzr}
\centering
\includegraphics[width=\textwidth]{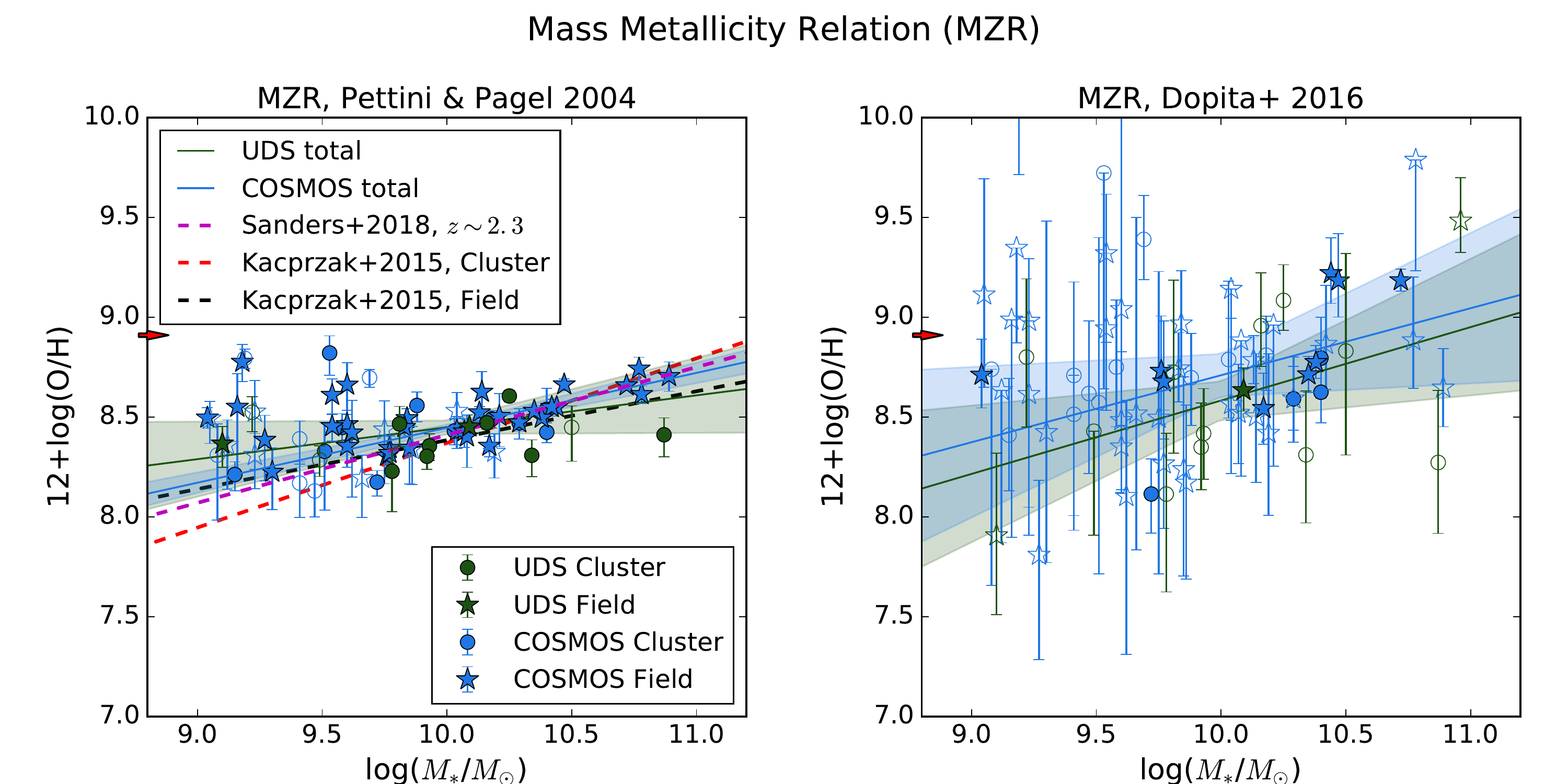}
\includegraphics[width=\textwidth]{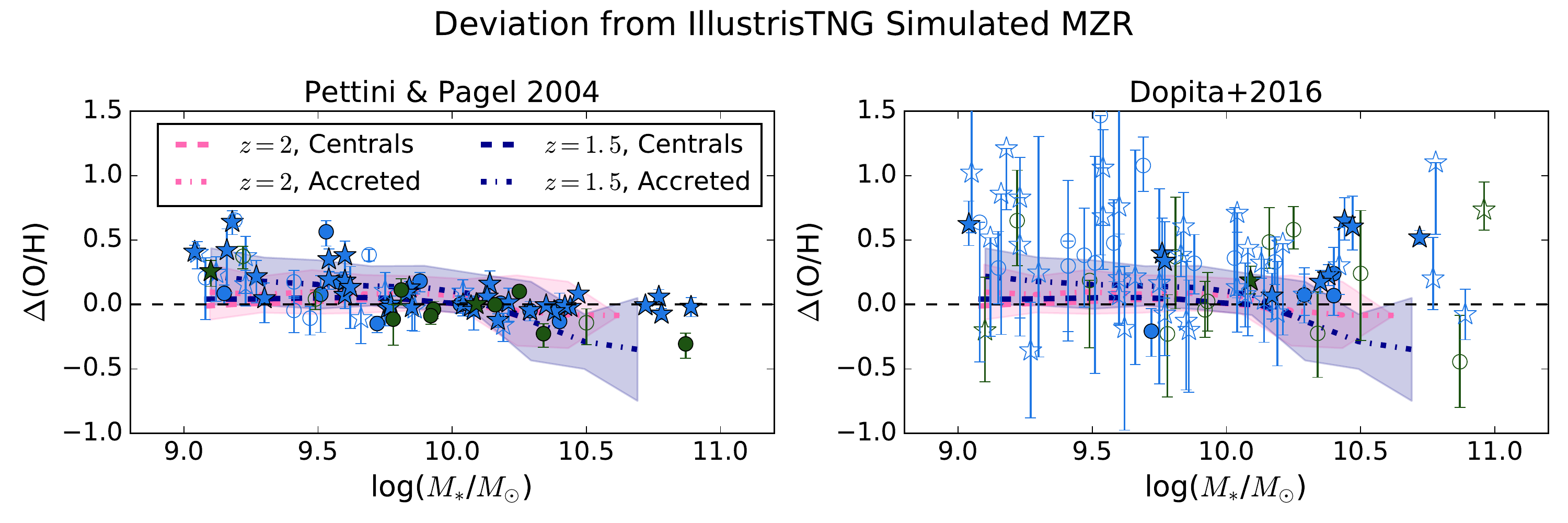}
	\caption{Top Left: Metallicity values determined by the indicator presented in \cite{Pettini2004}. 
	All emission lines are above a 2$\sigma$ detection limit. When at least one emission line shows significant sky interference, the point is unfilled. 
	We compare to the MOSDEF metallicity measurements \citep{Sanders2018} at $z\sim2.3$, and to the ZFIRE measurements of the COSMOS proto-cluster and field in \citet{Kacprzak2015}.
	Small red arrows on the left side of each plot mark the location of solar metallicity.
	Top Right: Metallicity values determined by the indicator presented in \citet{Dopita2016}.
	We show our best-fit linear relations for UDS and COSMOS in their respective colors. 
	Bottom rows: We plot the offset from the slope of the MZR in the IllustrisTNG simulation. The dotted and dashed lines in pink and blue are theoretical measures of the MZR from the Illustris simulation \citep{Gupta2018, Torrey2018}. 
	Centrals are galaxies in the center of their own dark matter potential (equivalent to a field sample), and accreted galaxies are equivalent to a cluster sample.
	We plot the deviation from the MZR predicted in IllustrisTNG with respect to $z=2$ centrals (black dashed line).}
\end{figure*}

\subsection{Stacking Emission Lines}
\label{sec:stacking}
As many of our galaxies have very faint emission lines (i.e. \NII, \hbeta, \SII), we stack our galaxy spectra into bins of stellar mass and environment to determine characteristics of their populations.
The stacked spectra include galaxies with emission line $SNR<2$, in contrast to our analysis with individual galaxies.
Spectra are collapsed over the same spatial aperture as determined in Section \ref{sec:emlines}, then normalized to the \halpha\ flux or \OIII\ flux, depending on the filter observed.
Objects are then interpolated onto a reference rest wavelength range preserving flux.
We take the median value of our stacks after rejecting outlier pixels with values greater than the median $\pm3$ NMAD (Normalized Median Absolute Deviation).
We perform our Gaussian line-fitting procedure from Section \ref{sec:emlines} to determine stacked fluxes.

\begin{figure*}[t] \label{fig:mzr_compareindicators}
\centering
\includegraphics[width=\textwidth]{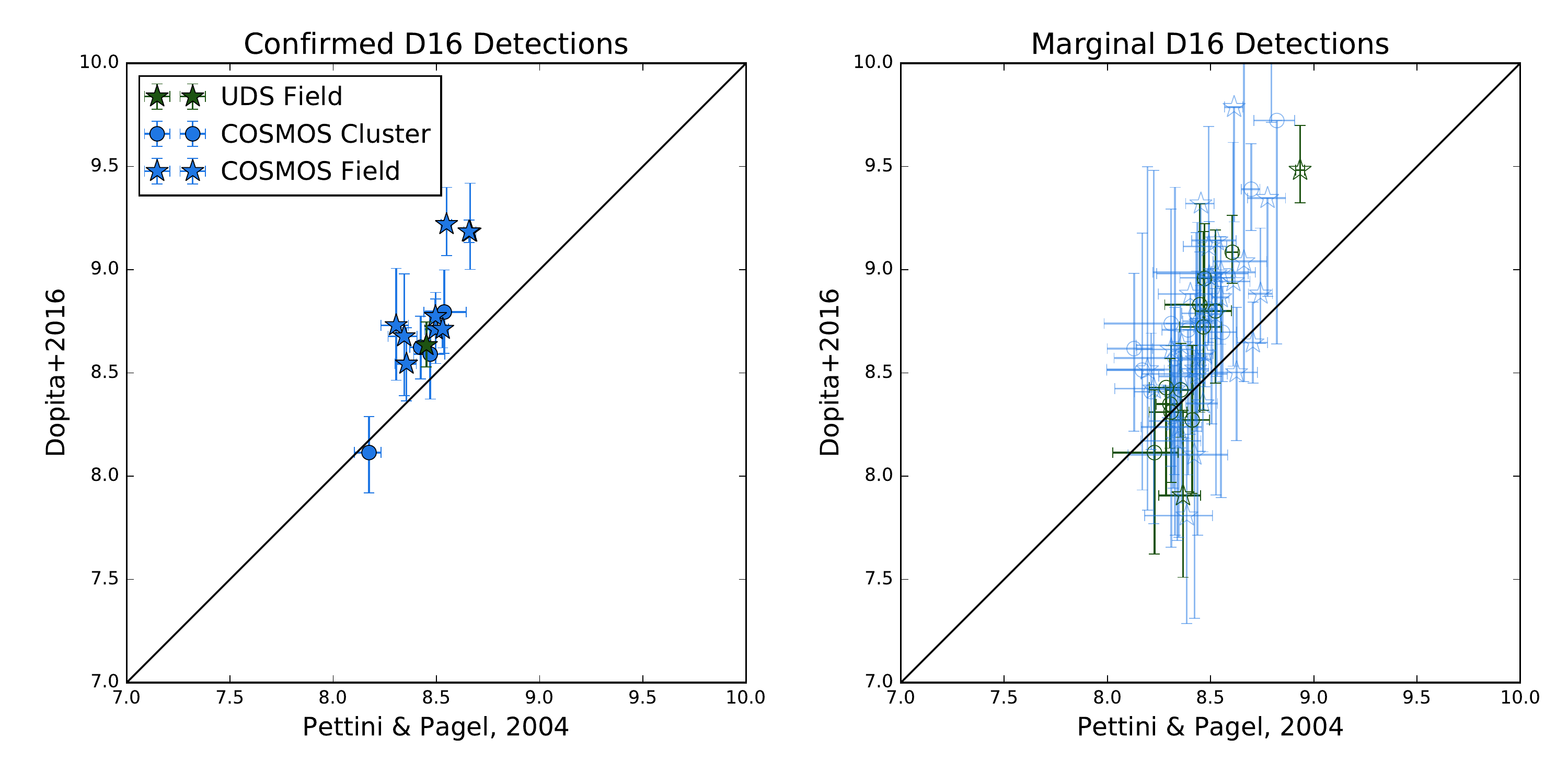}
	\caption{The direct comparison between metallicity indicators D16 \citep{Dopita2016} and PP04 \citep{Pettini2004}. Colors are as in Figure 2. Left: Detections where \halpha, \NII, and \SII\ emission lines are all free of significant sky interference. The one-to-one line, where measurements are equal, is the solid black line. Right: Detections where at least one emission line has significant sky interference. COSMOS galaxies are made transparent so UDS galaxies can be easily viewed.}
\end{figure*}

To separate the effects of AGN, we bin the objects identified as AGN in \citet{Cowley2016}. 
AGN are displayed separately in our BPT diagrams for reference.
\cite{Cowley2016} identifies AGN in the ZFOURGE sample using radio, IR, UV, and X-ray data.
AGN can affect placement on the BPT diagram as the emission is from shocked gas or photoionization by the hard ionizing spectrum of the AGN, rather than gas photo-ionized by young stellar populations.

\begin{figure*}[t] \label{fig:mzr_stack}
\centering
\includegraphics[width=\textwidth]{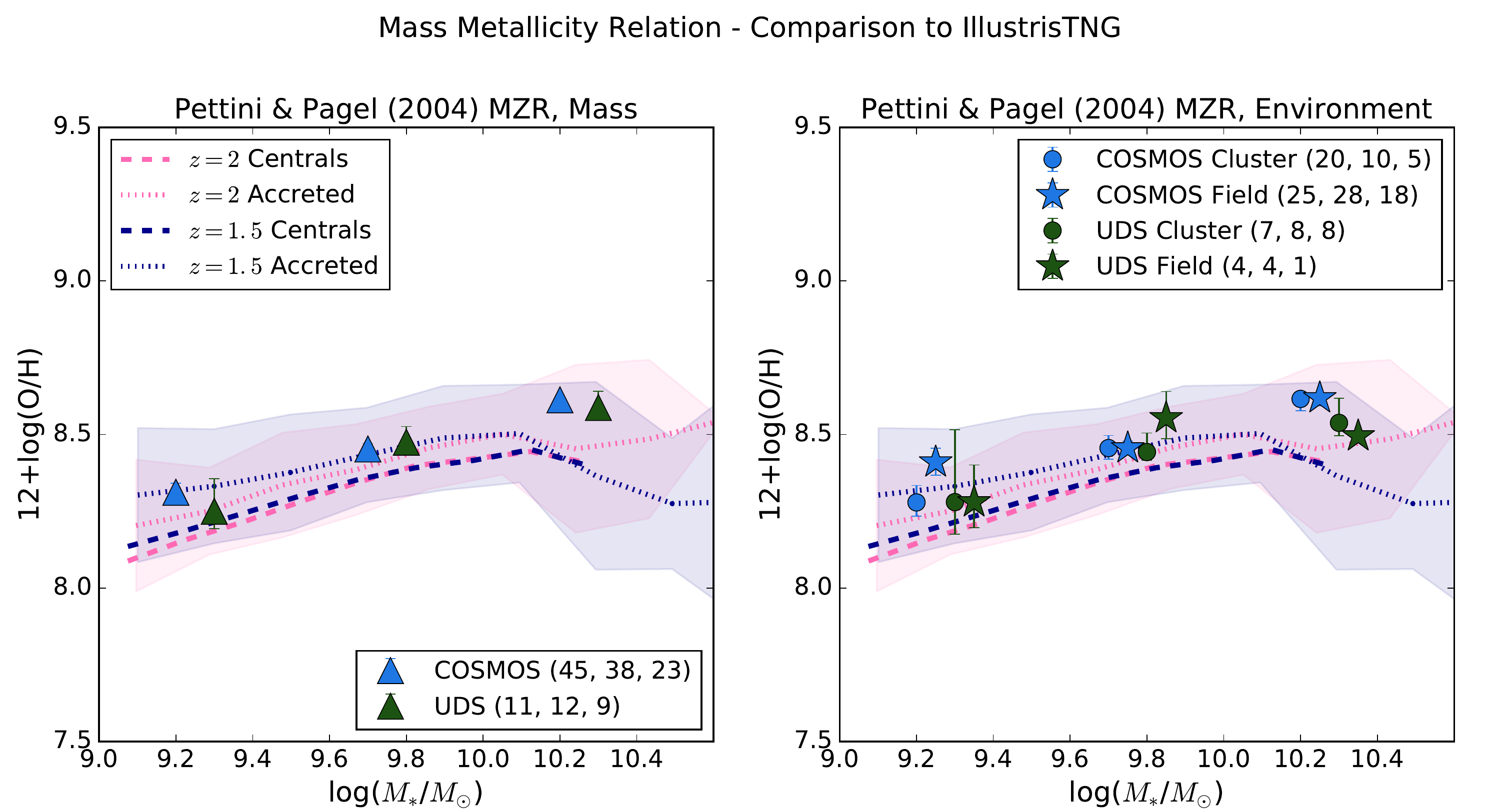}
\includegraphics[width=\textwidth]{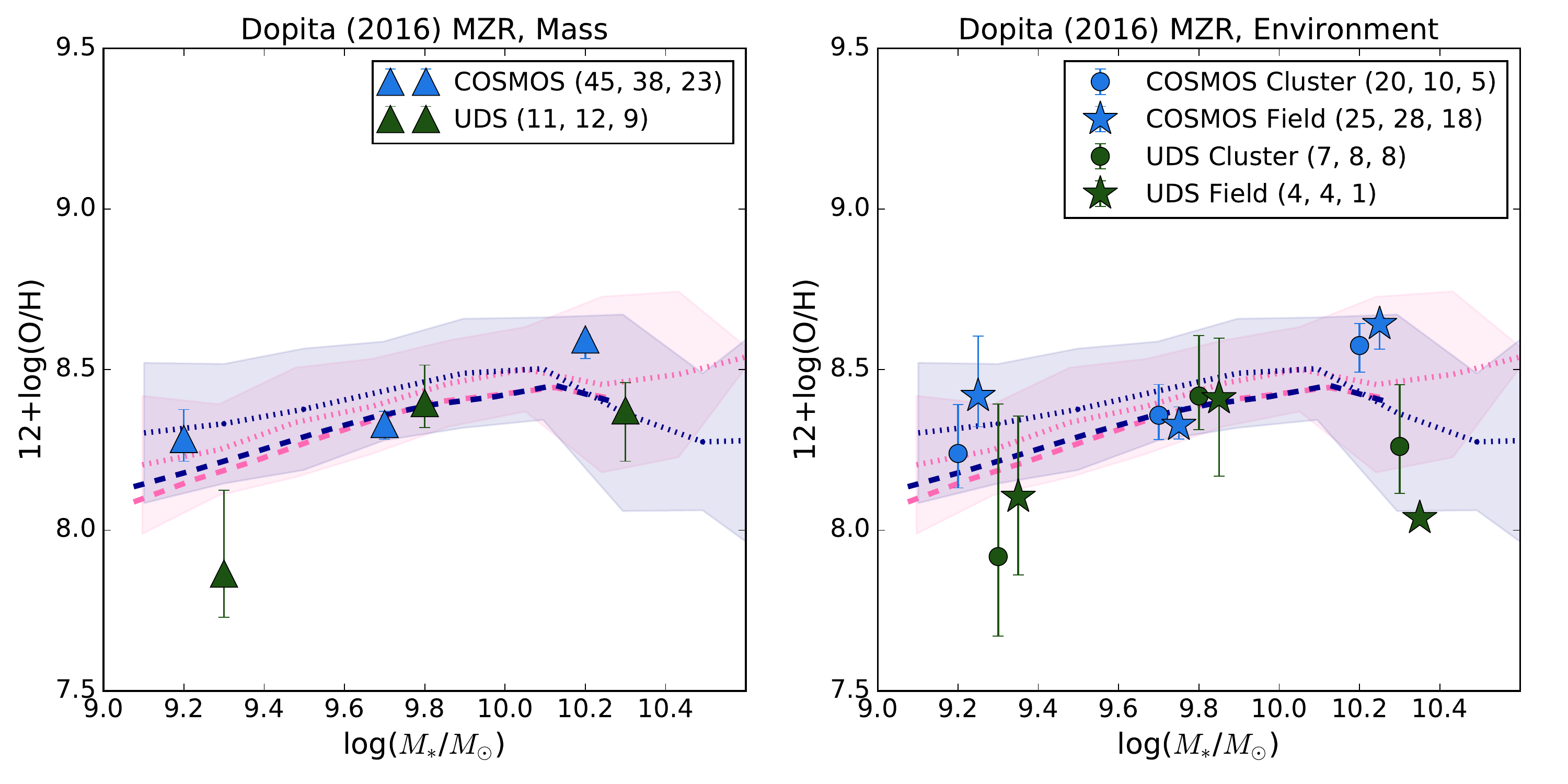}
	\caption{The stacked metallicity measurements from the sample. 
	Top rows are metallicities determined from the indicator presented in \citet{Pettini2004}, bottom rows are determined from the indicator in \citet{Dopita2016}. 
	Mass bins are offset in the \Mstar\ axis for clarity. 
	The IllustrisTNG simulations are shown in pink and blue, with scatter shown in shaded regions.
	Central galaxies are simulated galaxies that are central within their dark matter halos, equivalent to our field sample.
	Accreted galaxies are in-falling galaxies, equivalent to our cluster galaxy sample.
	Left: Metallicity values binned by mass. The low mass bin includes galaxies with log(\Mstar) $<9.54$, the medium mass bin from $9.54<$log(\Mstar)$<10.1$, and the high mass bin from log(\Mstar)$>10.1$. The legend displays the number of galaxies in low, medium, and high mass bins, respectively. Right: Metallicity binned by environment and mass.}
\end{figure*}

We stack our objects based on mass, environment, field (COSMOS vs. UDS), and detected AGN.
Objects are separated by i) low (\Mstar$<9.60$) vs. high (\Mstar$>6$) stellar mass objects, where \Mstar$=6$ is the median stellar mass of the COSMOS sample, and ii) objects identified within each proto-cluster (from the redshift limits of each proto-cluster) vs. objects that are field galaxies (See Sections \ref{sec:udssample} and \ref{sec:cosmossample}).

\section{Results}
\subsection{Gas-Phase Metallicity} \label{sec:metal}
\subsubsection{Comparing Strong-Line Indicators} \label{sec:indicatorcomp}
We apply the strong line diagnostic presented in \citet{Dopita2016} to determine gas-phase metallicities for a sample of our galaxies (Figure 2, right panels).
This indicator requires measurements of the \halpha, \NII, and \SII\ emission lines.
In the MOSFIRE K band, this indicator can be used for galaxies between $1.8<z<2.6$.
In the H band, it can be used for galaxies between $1.2<z<1.8$.
We can observe these lines for most of our objects in the same band on MOSFIRE, therefore absolute fluxes are not required.

However, the \SII\ doublet is quite faint ($\sim 10^{-18}$ergs/s/cm$^{2}/\AA$) for most objects in our sample, and in many of our objects we can only determine an upper bound of metallicity. The \SII\ doublet may only be slightly above the noise level, because of this we have introduced a signal to noise (S/N) cut of 2 on all emission lines.
\citet{Telford2016} showed that introducing S/N cuts can bias strong-line metallicity measurements depending on the choice of emission lines. 
They determined that S/N cuts on the \NII\ line will bias the sample against low-metallicity galaxies, but cuts on \halpha\ and \SII\ do not bias samples.
Therefore, we are biased toward high-metallicity galaxies, but this will be seen in both metallicity indicators we discuss, because both rely on measurements of \NII.

Additionally, in the UDS sample, the \SII\ 6731\AA\ emission line is in a region of high sky interference, so we mark most of our UDS objects as having unreliable \citet{Dopita2016} metallicities.
We mark objects with sky interference as unfilled points in Figure 2.
When we perform the same analysis using only objects with all emission lines in low sky regions, we find our COSMOS results are not significantly changed.
Only one UDS galaxy did not have any sky emission interference in the \SII\ lines, so we were not able to test this effect on the UDS population.

For comparison, we also determine metallicities from the PP04, also referred to as N2, indicator \citep{Pettini2004} (See Figure 2, left panels). 
We find that at lower values of metallicity $12+\log(O/H)<8.25$, the PP04 indicator predicts higher metallicity values than the D16 method, possibly indicating a floor on measurement sensitivity at low metallicity due to flux limitations on the faint \NII\ emission line (Figure 3).
The predicted D16 indicator typically is $\sim0.5$ dex higher than the PP04 indicator at low masses, $\log(M_{\ast}<9.5$.
Additionally, we compared our measurements to the MOSDEF sample  \citep{Sanders2018} and the previous measurements of the COSMOS proto-cluster and field \citet{Kacprzak2015}, which only used the PP04 indicator, and found similar results as our PP04 measurements (Figure 2, top panels).

Despite this observational difficulty, the strength of the D16 diagnostic is its independence from ionization parameter and ISM pressure, both of which have been shown to differ between local galaxy populations and those at high redshift \citep{Kewley2013,Steidel2014, Shapley2015, Kewley2015}.
Both the D16 and PP04 strong-line indicators depend on the correct calibration of the N/O ratio.
Current estimates of the N/O ratio at high-$z$ show identical values to local ratios, but at increased scatter \citep{Steidel2016, Maiolino2019}.

There is a large difference in scatter between the two metallicity indicators (Figure 3).
The NMAD scatter from the PP04 method (0.11 dex for COSMOS and 0.16 dex for UDS) is much lower than D16 (0.27 dex for COSMOS and 0.49 dex for UDS), most likely because there are fewer low-flux emission lines needed for the PP04 measurement.
\SII\ is a very faint emission line doublet, and for many of our galaxies we extracted flux limits rather than fluxes.
Additionally, the \SII\ line 6731\AA\ in the UDS proto-cluster is located in a high sky-noise region, making flux extraction difficult and we can only extract flux upper limits.
This likely explains the high scatter and super-solar metallicity values.

Because of sky emission that affects the extraction of \SII\ for the D16 indicator, we find that PP04 is the better indicator for gas-phase metallicity at \zsimtwo.
We emphasize that the D16 indicator is a powerful metallicity indicator and is relatively insensitive to the changing ionized gas properties of galaxies at high redshift.
Higher SNR emission lines of galaxies at \zsimtwo\ are needed to determine the gas-phase metallicities at the precision necessary to separate cluster and field populations.
Spectra from the upcoming thirty-meter class telescopes and space-based spectra (from e.g. JWST) will be valuable resources.

\subsubsection{Environment and Metallicity} \label{sec:environmentandmetallicity}

We find consistent results to the previous \zfire\ analysis of metallicity \citep{Kacprzak2015}, which determined that the COSMOS $z=2.095$ proto-cluster shared the same MZR as the field galaxies.
In \citet{Tran2015}, the UDS $z=1.62$ proto-cluster lacked evidence of a gas-phase metallicity offset from the field sample of \citet{Steidel2014}.
We find here that these proto-clusters also share the same MZR, and show no strong environmental influences.
Our results, given the scatter in our relations, are only sensitive to a metallicity difference of about 0.15 dex for the \citet{Pettini2004} indicator and 0.5 dex for the \citet{Dopita2016} indicator.

When we stack our galaxies by environment and stellar mass (see Figure 4), we find the similar results.
The MZR of proto-cluster and field populations are consistent within error, except in some very low-number bins (in the UDS population).
Like the theoretical values shown in the lower panels of Figure 2 at $z=1.5$ and $z=2$, we find no strong offsets between proto-cluster and field in gas-phase metallicity.

\subsubsection{Predictions from IllustrisTNG} \label{sec:illustristngintro}

We perform the first comparison of the gas-phase mass-metallicity relation as a function of environment to simulated results from IllustrisTNG.
IllustrisTNG determined gas-phase Mass-Metallicity Relations for $0<z<2$ \citep{Gupta2018, Torrey2018} and we find agreement with these predictions. 
The IllustrisTNG accreted/satellite galaxies were chosen from those living in halos of $10^{13} M_{\odot}/h$ and above at $z=0$, giving a total of 127 clusters.
The field/central sample was chosen from galaxies within their own dark matter halos of $10^{12.0} M_{\odot}/h$ at $z=0$, and with SFR$>$0.
Additionally, a stellar mass limit of $10^9 - 10^{10.5} M_{\odot}/h$ was imposed on all galaxies in the simulated sample.

\citet{Gupta2018} found that IllustrisTNG galaxies in clusters (accreted galaxies) at $z<1$ have an MZR metallicity enhancement of 0.15-0.2 dex from field galaxies (central galaxies) at the same redshifts.
At $z=1.5$ the predicted enhancement is only 0.05 dex, and at $z=2$ there is no enhancement.
At all redshifts, \citet{Gupta2018} reports a scatter in the MZR of 0.2 dex.
In the lower panels of Figure 2, we show the metallicity offset from the IllustrisTNG MZR relation for central galaxies (central galaxies within their dark matter halo, similar to our field galaxies) and accreted galaxies (galaxies within a shared dark matter halo accreted galaxies (in-falling galaxies in a dark matter halo belonging to another galaxy, equivalent to our proto-cluster galaxy samples) at $z=2$.
We can confirm that in both populations at $z=1.62$ and $z=2.095$, proto-cluster galaxies do not have significantly enhanced metallicities (Figure 2, lower panels, Figure 4).

The scatter of the IllustrisTNG simulated MZR remains consistent at all environments and redshifts, at a level of $0.2$ dex.
The scatter in our observed MZR values is only slightly smaller than the predicted scatter of the PP04 indicator in IllustrisTNG ($\sim0.1-0.15$ dex) but larger when using the D16 indicator ($\sim0.3-0.5$ dex).
Therefore we are unlikely to detect the minor ($<0.05$ dex) MZR enhancement seen at $z>1.5$ in the IllustrisTNG cluster galaxies.

\begin{figure}[h]  \label{fig:bpt}
	\centering
	\epsscale{1.25}
	\plotone{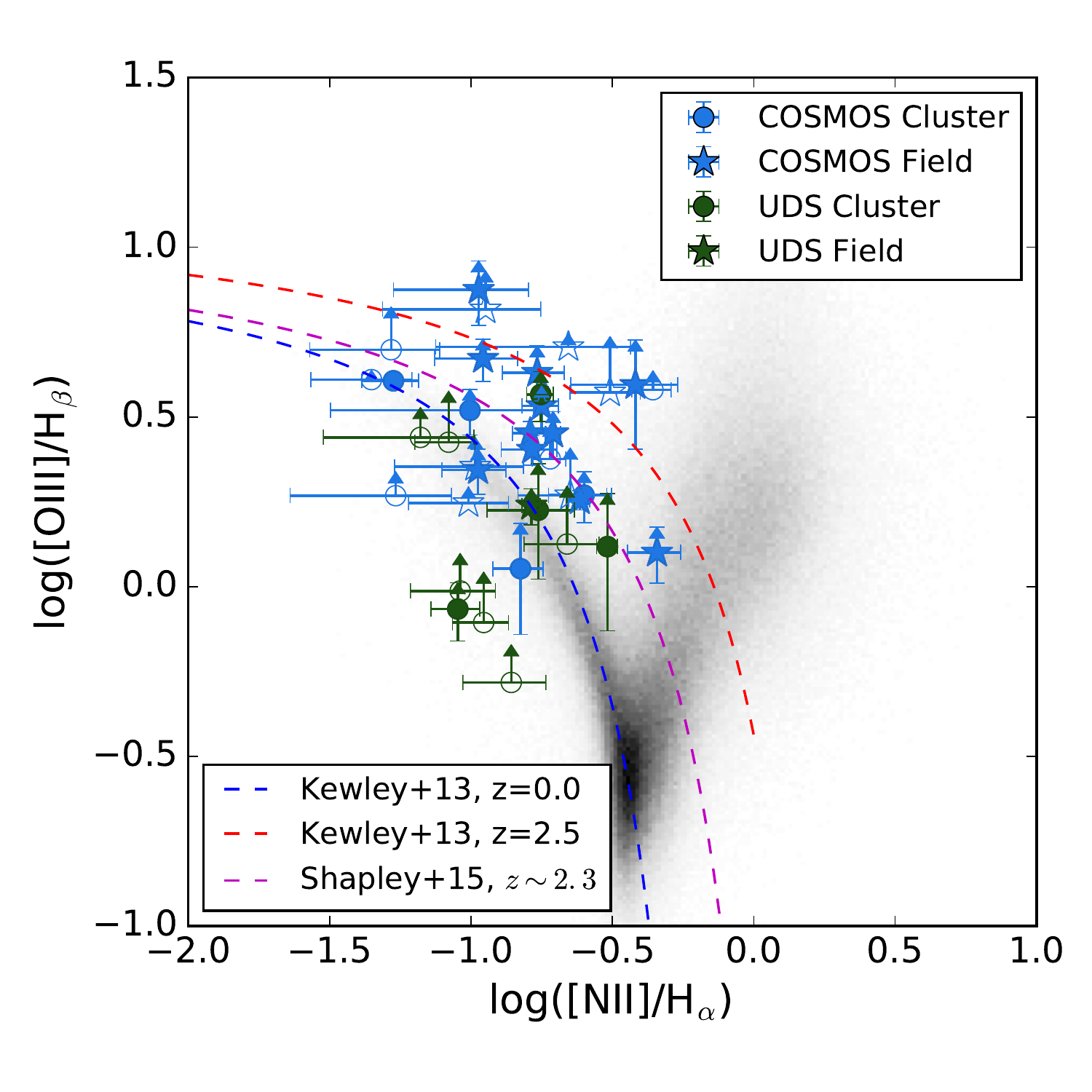}
	\caption{BPT diagram of galaxies within our two fields (COSMOS and UDS). Galaxies with all four emission lines \halpha, \NII, \hbeta, and \OIII\ with greater than 2$\sigma$ detection limit are shown. Galaxies with one or more emission lines with heavy sky interference are unfilled points. Colors and marker shapes are the same as in Figure 2. Shaded regions are SDSS galaxies. Colored curves from \citet{Kewley2013} are the upper-limit to the theoretical evolution of star-forming galaxies at $z=0$ and $z=2.5$.}
\end{figure}

\subsection{BPT Diagram} \label{sec:bptsection}

A subset of our galaxies in both proto-clusters have measurements of all emission lines necessary to apply Baldwin-Phillips-Terlevich (BPT) Diagnostics \citep{Baldwin1981}. 
These emission lines are \hbeta\ 4861\AA, \OIII\ 5007\AA, \halpha\ 6563\AA, and \NII\ 6585\AA.
We require our detections for all lines to be at SNR$>2$.

UDS galaxies appear to be offset from the COSMOS sample in the BPT diagram (Figure 5).
However the \hbeta\ line in UDS is located in a high sky-noise region, so our study may be biased toward objects with higher \hbeta\ flux.
A two-population Kolmogorov - Smirnov test confirms that, at a level of 0.005, that the COSMOS sample has a higher $\log$(\OIII/\hbeta) ratio than the UDS sample (at a p-value of 0.0013).
Deeper observations may be necessary to determine if sky interference is biasing this analysis, however our BPT KS test results hold when we reject objects where any emission line displays significant sky interference (Figure 5).

\begin{figure*}[h] \label{fig:bptstacks}
\centering
\includegraphics[width=\textwidth]{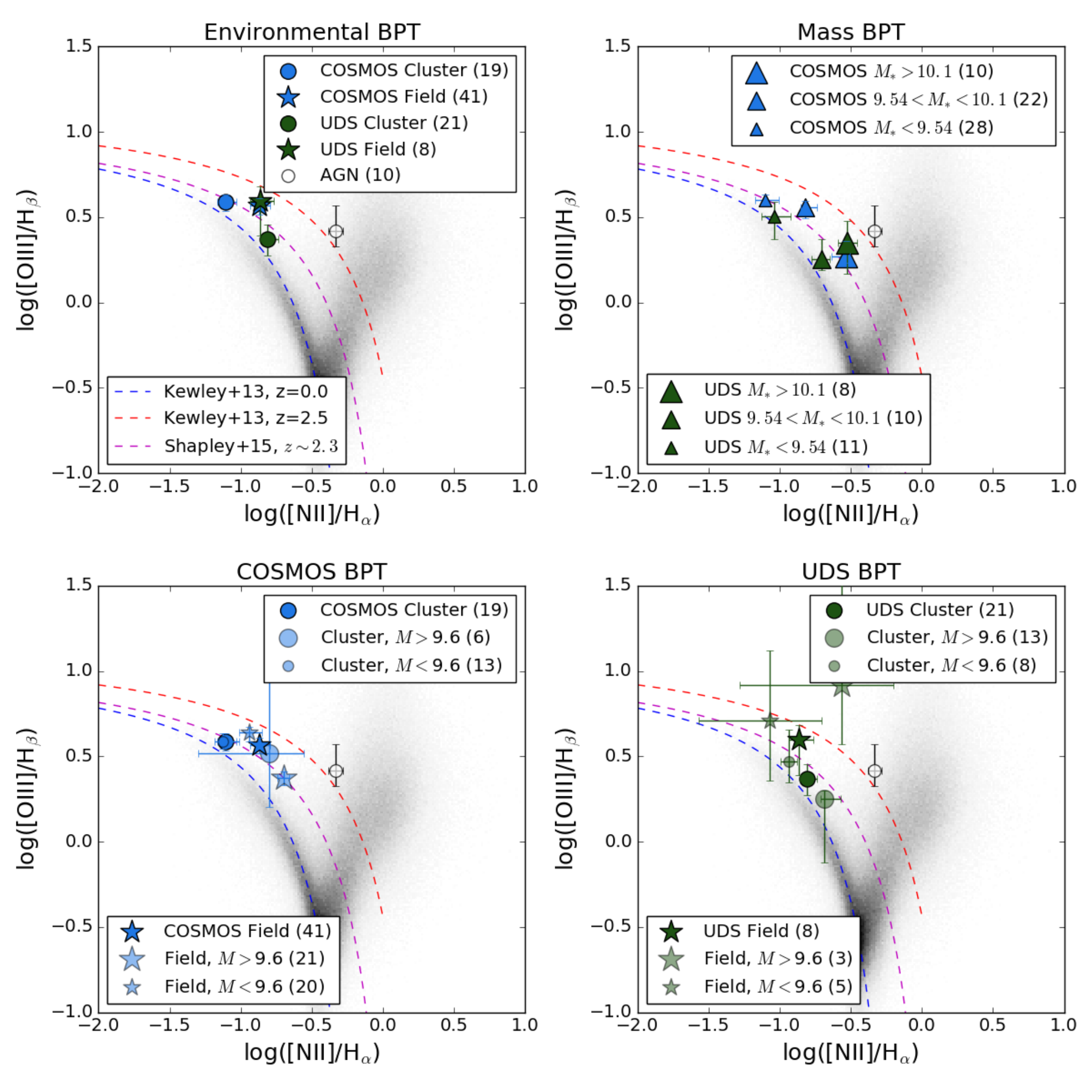}
	\caption{BPT diagram of galaxies within our two proto-clusters. We stack our galaxies according to their mass, environmental density, field (COSMOS or UDS), and inferred excitation mechanism (AGN identified via radio, IR, UV, or BPT diagram are separated from all stacks). Upper left: Stacking according to field and environmental density, where we find an offset in the \NII/\halpha\ ratio in COSMOS, and the \OIII/\hbeta\ ratio in UDS. Upper right: Stacking according to FAST-inferred stellar mass. We find a stronger effect from stellar mass in both fields. Lower left: The COSMOS $z\sim 2$ sample, split into low and high mass bins in both proto-cluster and field. The low and high mass bins are split at log(\Mstar) $=9.60$, which is the median of the COSMOS stacked sample. Lower right: The UDS sample similarly separated into high and low stellar mass bins with respect to environment.}
\end{figure*}

ISM conditions are known to change in high-redshift SFGs compared to local SFGs \citep{Kewley2013, Kewley2015, Shapley2015}.
Evolution of the ionization parameter (the amount of ionization a luminous source can produce in an HII region) does not evolve significantly between $0<z<1$ \citep{Paalvast2018}, but changes to more extreme ionizing conditions between $1<z<3$.
Additionally, changes in the $\log$(\OIII/\hbeta) ratio can be caused by changes in electron density, hardness of the ionizing field, and metallicity \citep{Kewley2015}.
Our results indicating an offset between the COSMOS and UDS proto-cluster show a tendency in the $z=2.095$ population to have more extreme ISM conditions than the $z=1.62$ population.
Additionally, both COSMOS and UDS offset from the SDSS sample in \OIII/\hbeta\ (Figure 5), consistent with an evolution in ionization parameter from local galaxies.

\subsubsection{Stacked BPT Measurements} \label{sec:bptstackedresults}

Our stacked results tell a similar story (Figure 6).
Galaxies are binned according to stellar mass (inferred from FAST), environment, AGN status (confirmed AGN are excluded from the sample but shown Figure 6), and subdivided into low and high stellar mass bins for each proto-cluster and field sample.
Spectra within a bin are then stacked.
In each case, the stellar mass has a strong influence on the position of stacked sample in the BPT diagram.

Higher stellar mass is known to correlate with higher metallicity and therefore higher \NII/\halpha.
We can confirm this correlation with the stacked spectra in the upper right panel of Figure 6.
Very clearly there is a sequence of increasing $\log$(\NII/\halpha) ratio with stellar mass in both the COSMOS (blue triangles) and UDS (green triangles) samples.

\subsubsection{UDS ($z=1.6$) vs. COSMOS ($z=2.1$)} \label{udsvcosmos}
In the upper left panel of Figure 6, we see galaxies in COSMOS and UDS binned purely by cluster status.
The COSMOS proto-cluster (blue solid circle) is offset from the field (blue solid star) by $\sim0.35$ dex in $\log$(\NII/\halpha), and no significant $\log$(\OIII/\hbeta) offset.
The $\log$(\NII/\halpha) offset is likely due to the difference in median mass between cluster and field populations.
The median mass of proto-cluster galaxies in COSMOS is $\log$(\Mstar)$=9.46$, and the median mass of the field galaxies is $\log$(\Mstar)$=9.75$.
The UDS proto-cluster (green solid circle), in contrast, shows a $\sim0.25$ dex offset from the field (green solid star) in $\log$(\OIII/\hbeta) and no $\log$(\NII/\halpha) offset.
Binned and stacked AGN are shown for reference (outlined black circle).
Both proto-cluster samples are located in the same $\log$(\OIII/\hbeta) range as the SDSS sample, and both field samples have consistent $\log$(\OIII/\hbeta) values as the \citet{Shapley2015} MOSDEF sample of $z\sim2.3$ field galaxies (fuchsia dashed line).

\subsubsection{Environment vs. Mass}
In order to disentangle the effects of environment and mass, we separate our proto-cluster and field sample by stellar mass in the bottom two panels of Figure 6.
Our proto-cluster and field bins are binned into a low mass (log(\Mstar) $<9.60$, transparent small points) and high mass (log(\Mstar) $>9.60$, transparent large points) bin.
The mass bins were chosen because it is the median mass of the stacked COSMOS sample.

We find no significant differences (within error) in the $\log$(\OIII/\hbeta) ratio within low mass bins in both COSMOS and UDS.
In UDS, this is a 0.24 dex offset, and in COSMOS 0.05 dex.
In the COSMOS low mass bin, there is a 0.18 dex offset in the $\log$(\NII/\halpha) ratio, at a 1$\sigma$ significance.

In the high mass bins, we find tentative offsets in the $\log$(\OIII/\hbeta) ratio at the level of 0.67 dex in UDS, and 0.15 dex in COSMOS.
However, the significance of the difference is subject to small number statistics in the UDS field population, and the COSMOS cluster population.
When the low and high mass bins are stacked using a different mass cut-off, the offsets change.

To analyze the effects of stellar mass binning, we compare these results with separating low and high mass galaxies at $\log$(\Mstar)$=9.72$, the median mass of the total UDS and COSMOS populations, and $\log$(\Mstar)$=9.46$, the median mass of the stacked COSMOS cluster galaxies.
We find that at high mass cuts, and a lower number of galaxies in the high mass bins, the UDS low mass field bin displays a 0.40 dex (2$\sigma$) $\log$(\OIII/\hbeta) ratio offset from the proto-cluster stack. 
As the mass cuts become lower, the \OIII/\hbeta\ offset moves to within error (0.34 dex). 
In the high mass bins we see the opposite effect, where at a smaller bin size and higher mass cut, we see no significant offset (0.13 dex, within errors), and at a lower mass cut we see a 2$\sigma$ offset (0.68 dex). 

Due to low numbers and high sky noise levels in the \hbeta\ emission line in the UDS field population, we consistently derive high errors for the high mass field stacked flux ratios. 
Therefore, the significance of any offset would be dependent on adding future data to the field bins. 
We still include these galaxies for a complete analysis.

In the COSMOS dataset, we find that at high mass cuts, the low mass COSMOS sample has no significant \OIII/\hbeta\ offsets (0.05 dex), but at lower mass cuts we see a 0.11 dex offset.
The significance of this offset is questionable due to the high errors seen in the high mass proto-cluster stack.
The \NII/\halpha\ ratio displays the inverse correlation, at high mass cuts there is a 2$\sigma$ offset (0.25 dex), and no offset at low mass cuts. 
In the high mass COSMOS stacks, we see large errors in the cluster bin due to low counts, but a consistent offset in \NII/\halpha\ ratio between cluster and field of around 3$\sigma$ (0.22 dex).

We discuss the possible explanations for the \OIII/\hbeta\ offset, if confirmed by larger datasets, in Section \ref{sec:analysis}.
However, we emphasize the speculative nature of the \OIII/\hbeta\ ratio offset due to low bin counts and low number statistics. 
A larger sample of high mass objects in both UDS and COSMOS would provide a more robust analysis of the $\log$(\OIII/\hbeta) offset.

\section{Discussion} \label{sec:analysis}

The \zfire\ survey was conducted to determine if and when a galaxy's environmental density at high redshift plays a role in the evolution of its observed properties \citep{Tran2015, Alcorn2016}.
Multiple analyses focused on star-formation rate \citep{Koyama2013a, Koyama2013,Tran2015,Tran2017}, metallicity \citep{Kacprzak2015, Tran2015, Kacprzak2016}, ionized gas characteristics \citep{Kewley2016}, gas content \citep{Koyama2017, Darvish2018} and kinematics \citep{Alcorn2016, Alcorn2018} have shown that the proto-clusters observed at $z>1.5$ have no significant environmental trends.

\subsection{\OIII/\Hbeta\ Ratios and Environment}
Despite previous non-detections of environmental effects on galaxy properties in the COSMOS proto-cluster at $z\sim2.0.5$ \citep{Kewley2016}, we find a tentative environmental effect on galaxy \OIII/\hbeta\ ratio when we stack high mass galaxies (Section \ref{sec:stacking}).
However, depending on the mass cuts for the stacks, these offsets can disappear.
The \OIII/\hbeta\ ratio correlates with the ionization parameter, electron density, hardness of the ionizing field, and metallicity.
If confirmed at a significance of 2$\sigma$ with a larger dataset, this offset would suggest that dense environment might have an effect on  the ionized gas characteristics of a galaxy.
However, these results are poorly constrained due to large errors and small number statistics, so further observations and longer exposures of these galaxies are necessary to confirm these results.

The \OIII/\Hbeta\ offset between cluster and field is seen in both the UDS and COSMOS populations at high mass (Section \ref{sec:bptstackedresults}).
However, we found metallicity offsets were not found to be correlated with environment, agreeing with prior studies (Section \ref{sec:environmentandmetallicity}) \citep{Kacprzak2015, Tran2015}. 
Taking these two results together suggests field galaxies have a higher ionization parameter than proto-cluster galaxies in UDS, whereas proto-cluster galaxies in COSMOS have a higher ionization parameter than the field \citep{Kewley2015}.

Additional metallicity diagnostics such as $R_{23}$, which utilizes \OIII4959,5007 \AA\ and \Hbeta4861\AA\ fluxes, would be needed to confirm this result, and  rule-out the choice of strong-line metallicity indicator as cause of null detection in metallicity.
A change in the \OIII/\hbeta\ ratio would not affect PP04 and D16 indicators, which only use \halpha, \NII, and \SII\ fluxes.
In contrast, the $R_{23}$ indicator would be affected by a change in the \OIII/\hbeta\ ratio, as $R_{23}$ is dependent on measurements of \OIII\ and the \OII\ doublet fluxes.
Our data set includes \OII$3727,29 \AA$ fluxes for a limited number of objects, so $R_{23}$ measurements are not performed, but will be an important test of our results in the future.

\section{Conclusions}
\label{sec:conclusions}

We perform an analysis of the ionized gas and gas-phase metallicity of two over-dense regions: the UDS proto-cluster at $z=1.62$ and COSMOS proto-cluster at $z=2.095$.
The rest-frame optical diagnostic lines \hbeta, \OIII, \halpha, \NII, and \SII\ are measured using the NIR spectrograph on Keck, MOSFIRE.
Fluxes of these emission lines are extracted via a Gaussian fit to the 1D spectrum.

Absolute fluxes of \halpha, \NII, and \SII\ are used to calculate the gas-phase metallicity using the abundance indicator presented in \citet{Dopita2016}, and compare to the metallicities derived using the method of \citet{Pettini2004}. 
We find no gas-phase metallicity offsets between our two proto-clusters at differing redshifts, or any metallicity enhancement relative to field galaxies at the redshifts of these galaxies.
Our results are consistent with predictions from IllustrisTNG \citep{Torrey2018, Gupta2018}.
We compare the utility of the PP04 and D16 indicators, and find that low \SII\ fluxes and high sky interference limit the use of the D16 indicator for our $z=1.6$ and $z=2.1$ samples due to high scatter ($\sim0.5$ dex).

For the subset of galaxies where all four diagnostic lines are measured, we analyze the nebular gas properties using the BPT diagram.
We find an offset in the \OIII/\hbeta\ ratio, or the ionization parameter in each population from the local SDSS sample.
The offset is consistent with photoionization models at high redshift, due to increasingly extreme ISM conditions (e.g. higher ionization parameter, higher electron density, harder ionizing field) \citep{Kewley2013, Kewley2015}.
When objects are stacked, we find a tentative $1-2\sigma$ $\log$(\OIII/\hbeta) enhancement in high mass ($>\log$(\Mstar)$=9.60$) field galaxies in the UDS stacks compared to high mass proto-cluster galaxies.

Further observations of over-dense regions of galaxies at \zsimtwo\ are needed to determine the significance of our results. 
In particular, larger sample sizes are needed in low-sky emission redshift windows, or with longer integration times so that \SII\ emission line fluxes are better constrained. 
Additionally, more measurements of clusters and proto-clusters at $z\sim1.0 - 1.5$ are needed, where IllustrisTNG predicts a metallicity enhancement relative to the field.
Likewise, cosmological simulations need to predict how ionization parameter and nebular gas properties change with environment.

\acknowledgments

The authors would like to thank the anonymous referee for comments which helped improve this paper.
L. Alcorn thanks the LSSTC Data Science Fellowship Program, her time as a Fellow has benefited this work.
This work was supported by a NASA Keck PI Data Award administered by the NASA Exoplanet Science Institute. 
Data presented herein were obtained at the W. M. Keck Observatory from telescope time allocated to NASA through the agency's scientific partnership with the California Institute of Technology and the University of California. 
This work is supported by the National Science Foundation under Grant \#1410728.  
GGK acknowledges the support of the Australian Research Council through the Discovery Project DP170103470. 
CMSS acknowledges funding through the H2020 ERC Consolidator Grant 683184. 
We acknowledge the Mitchell family, particularly the late George P. Mitchell, for their continuing support of astronomy. 
The authors wish to recognize and acknowledge the very significant cultural role and reverence that the summit of Mauna Kea has always had within the indigenous Hawaiian community. 
We are most fortunate to have the opportunity to conduct observations from this mountain.

\bibliography{library}

\end{document}